# EffiFusion-GAN: Efficient Fusion Generative Adversarial Network for Speech Enhancement

Bin Wen
School of Computer Sciences
Universiti Sains Malaysia
Penang, Malaysia
wenbin@student.usm.my

Tien-Ping Tan
School of Computer Sciences
Universiti Sains Malaysia
*Penang, Malaysia*
tienping@usm.my

*Abstract*—We introduce EffiFusion-GAN (Efficient Fusion Generative Adversarial Network), a novel deep learning model designed to enhance speech processing tasks by leveraging advanced techniques. Our model incorporates three primary innovations. Firstly, we employ Depthwise Separable Convolutions within a Multi-Scale Convolutional Block to significantly reduce computational complexity while capturing rich features at multiple scales. This approach enhances the model's ability to process diverse auditory inputs efficiently. Secondly, the model features an enhanced attention mechanism that includes dual Layer Normalization and optimized residual connections, improving stability and convergence during training. Lastly, dynamic pruning is applied to the convolutional layers, which reduces the model size without compromising performance, making it ideal for deployment in resource-constrained environments. These innovations combined within a GAN framework, allow EffiFusion-GAN to achieve superior speech enhancement results, balancing high quality with computational efficiency. Experimental results show that our proposed EffiFusion-GAN achieves a PESQ of 3.45 on the public VoiceBank+DEMAND dataset and Our model achieves the highest PESQ score under the same parameter settings.

## I. Introduction

With the introduction of many deep learning-based speech enhancement methods, the mainstream speech enhancement approaches fall into two categories: time-frequency domain speech enhancement methods and time-domain speech enhancement methods. Time-domain speech enhancement methods [1-3] operate directly on the time series of the speech signal, typically utilizing the raw waveform or processing the signal in frame units. However, time-domain speech enhancement methods may be limited in handling complex frequency variations, as they cannot directly capture and analyze the signal's frequency components, especially when dealing with non-stationary signals, making this limitation more apparent. Time-frequency domain speech enhancement methods convert the speech signal into a time-frequency representation using short-time Fourier transform (STFT) [5], then process it within the time-frequency domain, and finally reconstruct the time-domain signal through inverse short-time Fourier transform (ISTFT). These methods leverage the sparsity of the speech signal in the frequency domain, effectively separating noise from speech components, simultaneously analyzing the temporal and frequency characteristics of the signal, making them more effective in handling signals with frequency variations over time. However, time-frequency domain methods may suffer from inaccuracies in phase recovery, leading to degraded speech enhancement quality. Typically, due to the wrapping and non-structural characteristics of the phase spectrum, directly enhancing the phase spectrum poses a significant challenge, so it is often excluded. Nonetheless, recent studies have shown that phase information plays a crucial role in the perceptual quality of speech in SE methods, especially in low signal-to-noise ratio (SNR) [4] environments.

Recent speech enhancement research models [14,5,9] have predominantly used standard convolution. Standard convolution is often used in speech enhancement but suffers from high computational cost and large parameter sizes, especially for high-dimensional data. This increases the computational cost of the model [15,18] and may lead to overfitting issues [16,21]. With the advent of depthwise separable convolution [6-8], the problem of High computational cost and large parameter size has gradually been addressed. Depthwise separable convolution is an optimized convolutional computation method that decomposes standard convolution into two steps: depthwise convolution and pointwise convolution. It significantly reduces computational complexity. Depthwise convolution operates within each channel, while pointwise convolution performs a 1x1 convolution on the result of depthwise convolution. This approach drastically reduces the number of convolution operations required. However, since depthwise separable convolution splits the convolution operation into depthwise and pointwise steps, the features captured in each step may be fewer. Compared to standard convolution, it might miss some complex cross-channel features and result in reduced model robustness, particularly in tasks requiring high feature extraction accuracy, where the model's performance may not be as good as standard convolution. Therefore, to improve the effectiveness of speech enhancement while reducing parameters, it is crucial to ensure that the model's robustness is not compromised while enhancing performance. Based on the discussions above and other scholars' research findings [10], we further explored the topic, proposing EffiFusion-GAN (Efficient Fusion Generative Adversarial Network).

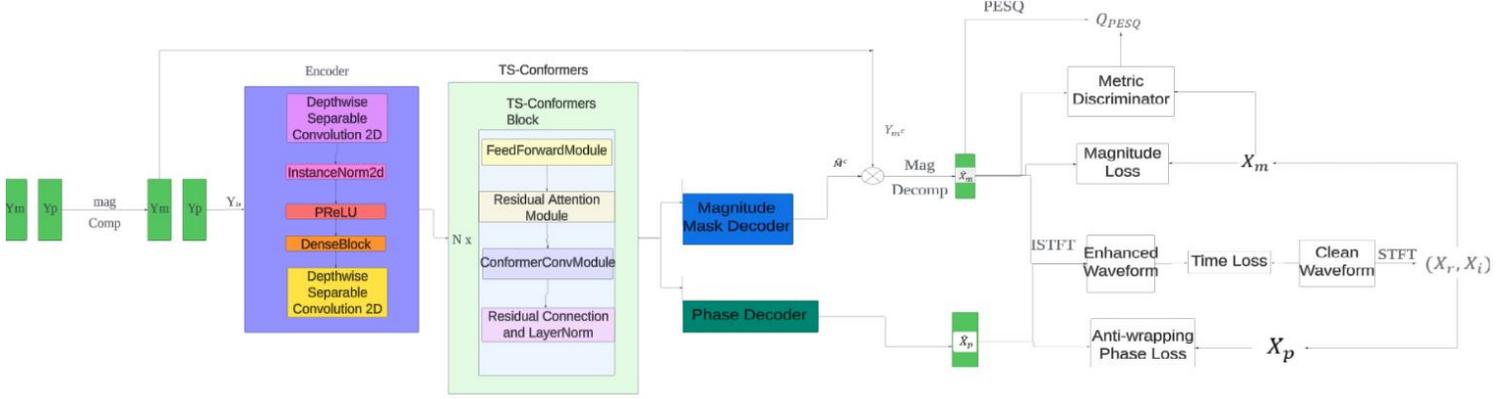

Fig. 1. An overview of the proposed EffiFusion-GAN architecture.

Our study has three contributions:

- Our model employs depthwise separable convolutions, significantly reducing the number of parameters and computational cost. This allows for efficient feature extraction from the noisy speech data while maintaining high performance.

- We incorporate residual connections combined with attention mechanisms in the conformer blocks. This design enhances the model's ability to capture both local and global dependencies, ensuring robust performance in noisy environments.

- The model utilizes L1 unstructured pruning, which removes less significant weights during training. This reduces the model's complexity and improves its computational efficiency without compromising speech enhancement quality.

## II. RELATED WORKS

Several important models have made significant contributions to the field of speech enhancement. SEGAN[15] utilizes a GAN architecture where the generator produces enhanced speech and the discriminator evaluates its quality. However, it has high computational costs and struggles with generalization to unseen noise. MetricGAN[16] optimizes perceptual metrics like PESQ instead of traditional loss functions, improving speech quality but limiting flexibility due to its reliance on predefined metrics. MetricGAN+[14] further optimizes perceptual metrics but still faces challenges related to metric dependency. DPT-FSNet[17] employs a dual-path transformer network to enhance both full-band and sub-band signals, achieving efficient enhancement with fewer parameters, although its PESQ performance is lower compared to models with larger parameters. CMGAN[9] combines the Conformer architecture with GANs to capture both local and global features in speech signals, significantly improving speech quality but at the cost of increased computational complexity. MP-SENet[13] introduces a parallel processing architecture for magnitude and phase spectra, offering superior PESQ scores, but its large parameter size (2.05M) makes it less efficient compared to models like DPT-FSNet.

## III. METHODOLOGY

The proposed model utilizes a specialized encoder-decoder architecture designed to denoise noisy speech and restore clear speech signals in the time-frequency domain. Initially, the noisy speech input undergoes a Short-Time Fourier Transform (STFT), extracting the magnitude spectrum ($Y_m$) and phase spectrum ($Y_p$), where F and T represent the total number of frequencies and frames, respectively. To improve the accuracy of magnitude mask prediction, we apply power-law compression to the magnitude spectrum $Y_m$ and stack it with the phase spectrum $Y_p$, forming the input feature $Y_{in}$. The encoder employs depthwise separable convolution layers to compress these features ($Y_{in}$) into time-frequency domain representations, effectively reducing computational complexity while maintaining efficient feature extraction. Subsequently, the model processes these compressed time-frequency domains using multiple convolution-enhanced transformers (TS-Conformers) with residual connections and multi-head attention mechanisms, which effectively capture dependencies across time and frequency, ensuring the preservation and enhancement of global features. During training, the model also prunes the convolutional layers, removing approximately 30% of redundant parameters, thereby reducing computational overhead and enhancing the model's generalization ability. Lastly, the magnitude mask and phase decoders operate concurrently to estimate the clean magnitude and phase spectra from the time-frequency domain, which are subsequently used to regenerate the enhanced speech waveform through inverse STFT (Short-Time Fourier Transform).

### 3.1 Model structure

#### 3.1.1 Encoder

As illustrated in Figure 1, the encoder encodes the input feature $Y_{in}$ into a time-frequency domain representation, which is characterized by efficient feature extraction capabilities and reduced computational complexity through deep compression.

The encoder architecture is composed of an initial depthwise separable convolutional block, a dilated DenseNet[11] module, and a concluding depthwise separable convolutional block. Each of these blocks includes a 2D depthwise separable convolution layer, an instance normalization layer, and a parametric ReLU (PReLU)[12] activation function. By leveraging depthwise separable convolutions, the model efficiently extracts features while significantly reducing the parameter count.

The initial depthwise separable convolutional block increases the number of channels in the convolution layers, thus enhancing feature dimensions and offering richer input for deeper processing. The dilated DenseNet module utilizes four convolution layers with dilation rates of 1, 2, 4, and 8 to extend the receptive field along the temporal axis, effectively capturing long-range dependencies. By using dense connections, the outputs from all depthwise separable convolution layers are combined, forming a more compact feature representation, which helps alleviate the vanishing gradient issue and simultaneously boosts feature expressiveness.

The final depthwise separable convolutional block adjusts the stride of the convolutional layers to downsample the features, compressing the high-dimensional features into a more compact representation. This not only reduces computational overhead but also ensures the preservation of critical features, providing high-quality input for subsequent multi-head attention mechanisms.

*3.1.2 Decoder*

As illustrated in Figure 1, the decoder is designed with two parallel modules[13]: the magnitude mask decoder and the phase decoder. The magnitude mask decoder initially estimates a magnitude mask from the time-frequency representation and applies it to the noisy magnitude spectrum to derive the clean magnitude spectrum. Traditionally, the magnitude mask is defined as $M = \frac{X_m}{Y_m}$, where $X_m$ represents the magnitude spectrum of the clean waveform xxx. However, this mask is unbounded, and to avoid this issue in practical applications, a sigmoid function is commonly used to constrain the mask values within the range (0, 1). Nevertheless, since the actual mask range may extend beyond this interval, this constraint leads to a discrepancy between the predicted and actual masks. To address this issue, in this study, we applied power-law compression to both $X_m$ and $Y_m$. We used a compression factor (c) to limit the prediction range of the mask. In our experiments, the compression factor was set to 0.3 to enhance prediction precision. Additionally, to further improve the model's accuracy, a learnable sigmoid (LSigmoid)[14] function was introduced to predict the compressed magnitude mask.

The phase decoder estimates the clean phase spectrum directly from the time-frequency representation. The recovery of phase information typically faces challenges due to its unstructured nature and phase wrapping issues. To overcome these challenges, we employed a parallel phase estimation architecture. This architecture generates pseudo-real $\hat{X}_p^{(r)}$ and pseudo-imaginary $\hat{X}_p^{(i)}$ components through two parallel 2D convolutional layers, followed by the calculation of the clean phase spectrum $\hat{X}_p^{(r)}$ using the arctan function with two arguments. This design not only enhances the reconstruction of phase information but also reduces the negative impact of phase discontinuities on speech quality, thereby effectively mitigating noise interference in the speech signal while preserving its naturalness.

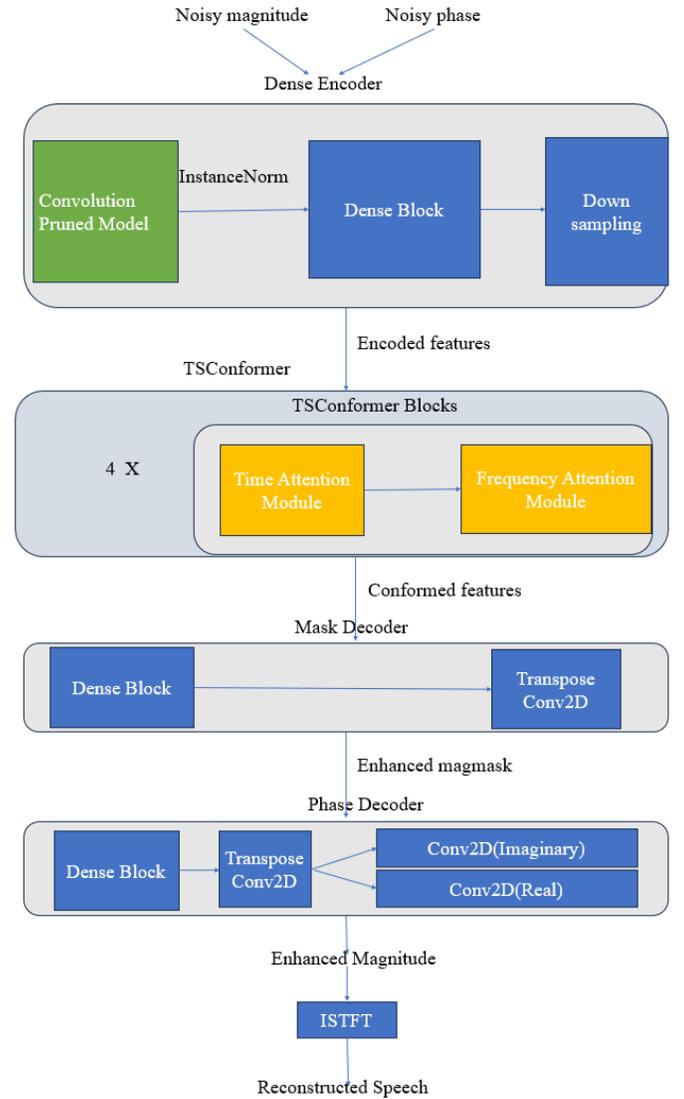

Fig. 2. The overall framework of Generator

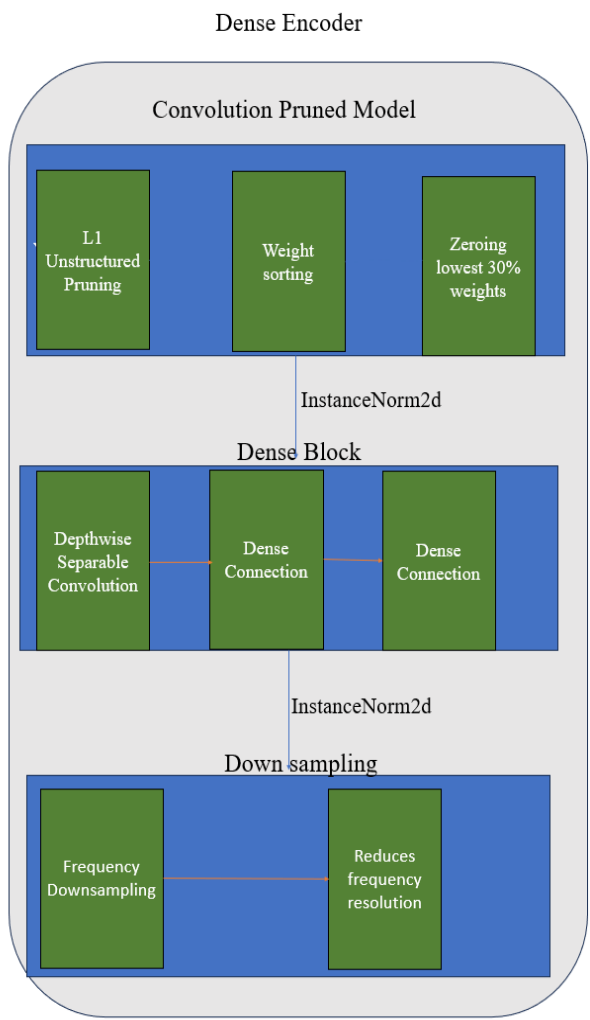

Fig. 3.   The details of the Dense Encoder

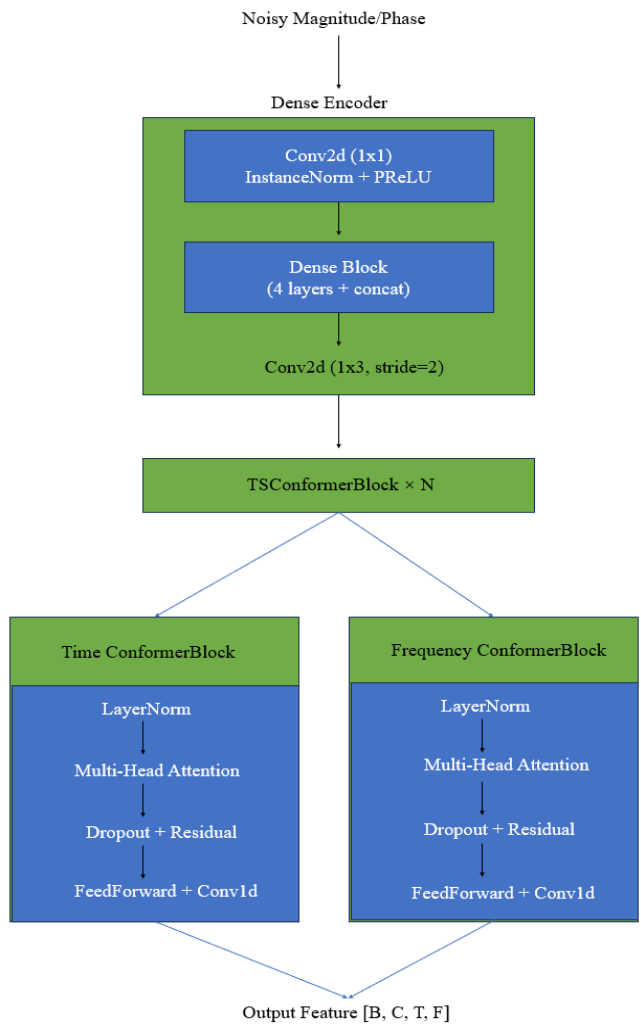

Fig. 4.   The figure of MHA and ResNet

### 3.2 Discriminator

The discriminator is a crucial component of the model, responsible for evaluating the similarity between the generated speech and the target clean speech. Its primary function is to differentiate between real and enhanced speech and provide a realism score that reflects how closely the enhanced speech matches the clean reference. It takes as input the magnitude spectra of both the clean and enhanced speech and processes them through multiple layers of feature extraction and transformation to make a final assessment. Figure 5 is the overall framework of Discriminator.

The discriminator first receives two inputs: the magnitude spectrum of the clean speech from the dataset and the magnitude spectrum of the enhanced speech generated by the model. These two inputs are concatenated along the channel dimension to form a two-channel feature map. This concatenation allows the model to process both signals together, facilitating a direct comparison between the clean and enhanced speech. The resulting feature map is structured as a four-dimensional tensor with the shape [B,2,T,F], where B is the batch size, T is the number of time frames, and F represents the frequency bins.

The concatenated feature map is then passed through a multi-layer convolutional network, which progressively extracts hierarchical representations while compressing both the time and frequency dimensions. The first convolutional layer applies a two-dimensional convolution (Conv2D) to extract local patterns in both time and frequency, capturing fundamental structures such as formants, harmonics, and spectral transitions. To efficiently process long sequences, a downsampling operation is applied, reducing both the time and frequency resolution while retaining key speech characteristics. This step allows the model to focus on essential features while reducing computational complexity.

To ensure stable training and efficient learning, the discriminator employs multiple normalization techniques. Spectral Normalization is applied to regulate the weight distribution of convolutional layers, preventing excessive weight growth and stabilizing the adversarial training process. Additionally, Instance Normalization (InstanceNorm2D) is used to normalize feature distributions independently for each sample, improving generalization across different speakers and recording conditions. To introduce non-linearity and enhance feature extraction, the Parametric ReLU (PReLU) activation function is applied, allowing the model to learn adaptable activation thresholds that better preserve information from both positive and negative activations.

After multiple layers of convolution and down sampling, the feature map retains high-level information but remains high-dimensional. To further compact this representation, the model applies Adaptive Max Pooling (AdaptiveMaxPooling2D), which compresses the feature map into a fixed-size representation. The advantage of adaptive pooling is that it produces a standardized output size regardless of the input dimensions, ensuring robustness to variations in speech length or spectral resolution. This step allows the model to handle speech signals of different durations while maintaining a fixed output structure.

The pooled feature map is then flattened into a one-dimensional vector and passed through a sequence of fully connected layers that progressively condense the information. The first linear transformation reduces the feature dimensionality while preserving the key patterns necessary for speech discrimination. To prevent overfitting and improve generalization, Dropout is applied between the linear layers, deactivating a portion of neurons during training. A second linear transformation further compresses the feature representation into a single scalar value, which serves as the realism score of the enhanced speech.

The final scalar output is mapped to a probability range of [0,1] using a learnable sigmoid function. This value represents the discriminator's confidence in the realism of the generated speech. A value close to 0 indicates that the enhanced speech significantly deviates from clean speech, suggesting the presence of residual noise, distortions, or unnatural artifacts. Conversely, a value close to 1 signifies that the enhanced speech closely resembles clean speech, meaning that the noise has been effectively removed while preserving natural speech characteristics.

When training the model, we employed a series of meticulously designed loss functions and optimization strategies to ensure the model's superior performance in the speech enhancement task. Firstly, the model utilizes a time-domain loss (Time Loss), which ensures temporal accuracy by computing the L1 distance between the generated speech waveform $\hat{x}$ and the ground truth speech waveform x. The formula is given by:

$$L_{Time} = ||x - \hat{x}||_1 \quad (1)$$

Secondly, the model employs a Magnitude Loss, which optimizes the spectral information of the generated speech by calculating the Mean Squared Error (MSE) between the ground truth magnitude spectrum $X_m$ and the generated magnitude spectrum $\hat{X}_m$. The formula is given by:

$$L_{Mag} = ||X_m - \hat{X}_m||_2^2 \quad (2)$$

Additionally, the model introduces a Complex Loss, which enhances the model's reconstruction capability in the complex domain by calculating the Mean Squared Error (MSE) separately for the real and imaginary parts of the complex spectrum. $X_r$ and $X_i$ represent the real and imaginary parts of the ground truth complex spectrum, respectively, while $\hat{X}_r$ and $\hat{X}_i$ denote the real and imaginary parts of the predicted complex spectrum generated by the model. The formula is given by:

$$L_{Mag} = ||X_r - \hat{X}_r||_2^2 + ||X_i - \hat{X}_i||_2^2 \quad (3)$$

To further enhance the quality of the generated speech, we designed a comprehensive Phase Loss, which includes the Instantaneous Phase Loss (LIP) that measures the difference between the actual phase $X_p$ and the predicted phase $\hat{X}_p$ the Group Delay Loss (LGD) that optimizes the temporal consistency by focusing on the group delay characteristics of phase changes over time, and the Instantaneous Angular Frequency Loss (LIAF) that ensures the smoothness and continuity of the phase across the frequency domain. These components work together to address phase-wrapping effects, ensuring the accuracy of the generated phase spectrum. The formula is given by:

$$L_{Pha} = L_{IP} + L_{GD} + L_{IAF} \qquad (4)$$

The Perceptual Loss (Metric Loss) evaluates the perceptual quality of the generated speech through a discriminator, ensuring that the final output speech signal is more natural and realistic in terms of subjective listening experience. The formula is given by:

$$L_{Metric} = F \times mse_{loss}(metric_g \times flatten(), one_{labels}) \quad (5)$$

Finally, the weighted combination of all these loss functions constitutes the overall loss for the generator:

$$L_{Generator} = 0.05 \times L_{Metric} + 0.9 \times L_{Mag} + 0.3 \times L_{Pha} + 0.1 \times L_{Com} + 0.2 \times L_{Time} \qquad (6)$$

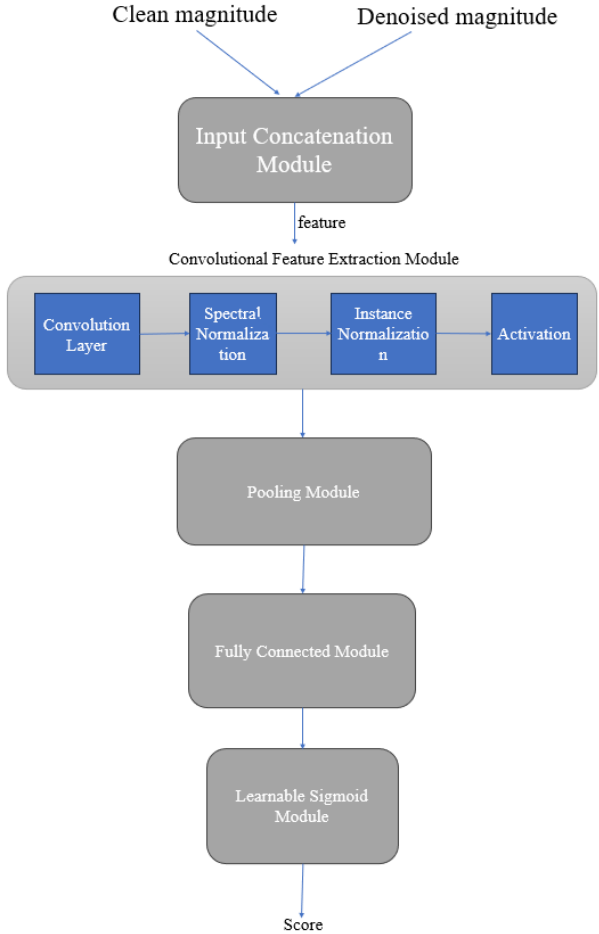

Fig. 5. The overall framework of Discriminator

## IV. EXPERIMENTS

### 4.1 Datasets and experimental setup

We carried out experiments using the publicly available VoiceBank+DEMAND dataset[19]. This dataset includes clean and noisy speech samples originally sampled at 48 kHz, which were resampled to 16 kHz for PESQ metric evaluation. The clean speech data is sourced from the Voice Bank corpus, comprising 11,572 audio clips from 28 speakers in the training set, and 872 clips from 2 unseen speakers in the test set. The noisy speech was created by combining clean audio with 10 different noise types (8 from the DEMAND database[20] and 2 artificially generated) across various Segmental Signal-to-Noise Ratios (SSNR). For testing, 5 unseen noise types were used with SNRs of 2.5 dB, 7.5 dB, 12.5 dB, and 17.5 dB

### 4.2 Comparison with other speech denoising Methods

We selected several representative time-domain speech enhancement methods (such as SEGAN, DEMUCS, and SE-Conformer) and time-frequency domain speech enhancement methods (including MetricGAN, MP-SENet, PHASEN, MetricGAN+, and four of the latest SOTA methods) to compare with EffiFusion-GAN. To evaluate the quality of the enhanced speech, we employed six commonly used objective evaluation metrics: PESQ, Segmental Signal-to-Noise Ratio (SSNR), Short-Time Objective Intelligibility (STOI), and three composite metrics (CSIG, CBAK, and COVL). For all these metrics, higher values indicate better performance. Based on Table I, although EffiFusion-GAN slightly underperforms MP-SENet in metrics such as PESQ, it utilizes only half as many parameters. When compared with DPT-FSNet, which has a similar parameter count, EffiFusion-GAN consistently outperforms it across all metrics. In summary, our proposed EffiFusion-GAN ensures high metric performance while minimizing the parameter footprint as much as possible.

### 4.3 Ablation study

To validate our design choices, we conducted an ablation study as shown in Table II. To investigate whether depthwise separable convolutions effectively reduce the model's parameter count, we compared the EffiFusion-GAN with standard convolutions. Additionally, we examined the impact of residual attention mechanisms and pruning on model performance by conducting comparative experiments. The results indicate that, while the model without depthwise separable convolutions showed some improvement in certain metrics, its parameter count nearly doubled. Moreover, removing the residual attention mechanism or pruning led to a noticeable decline in the model's overall performance.

TABLE I. Comparison with other methods on VoiceBank+DEMAND dataset. "-" denotes the result is not provided in the original paper.

| Method | Year | Input | Param | PESQ | CSIG | CBAK | COVL | SSNR | STOI |
|---|---|---|---|---|---|---|---|---|---|
| Noisy | - | - | - | 1.97 | 3.35 | 2.44 | 2.63 | 1.68 | 0.91 |
| SEGAN[15] | 2017 | Waveform | 43.18M | 2.16 | 3.48 | 2.94 | 2.80 | 7.73 | 0.92 |
| MetricGAN[16] | 2019 | Magnitude | - | 2.86 | 3.99 | 3.18 | 3.42 | - | - |
| MetricGAN+[14] | 2021 | Magnitude | - | 3.15 | 4.14 | 3.16 | 3.64 | - | **0.96** |
| DPT-FSNet[17] | 2021 | Complex | **0.88M** | 3.33 | 4.58 | 3.72 | 4.00 | - | **0.96** |
| CMGAN[9] | 2022 | Magnitude+Complex | 1.83M | 3.41 | 4.63 | 3.94 | 4.12 | 11.10 | **0.96** |
| PHASEN[18] | 2020 | Magnitude+Phase | - | 2.99 | 4.21 | 3.55 | 3.62 | 10.08 | - |
| MP-SENet[13] | 2023 | Magnitude+Phase | 2.05M | **3.50** | **4.73** | **3.95** | **4.22** | **10.64** | **0.96** |
| EffiFusion-GAN | 2024 | Magnitude+Phase | 1.08M | 3.45 | 4.71 | 3.91 | 4.18 | 10.12 | **0.96** |

TABLE II. Results of the ablation study

| Method | Param | PESQ | CSIG | CBAK | COVL |
|---|---|---|---|---|---|
| EffiFusion-GAN | 1.08M | 3.45 | 4.71 | 3.91 | 4.18 |
| w/o Depthwise | 2.04M | 3.47 | 4.76 | 3.90 | 4.17 |
| w/o Res | 1.08M | 3.35 | 4.64 | 3.86 | 4.08 |
| w/o pruning | 1.75M | 3.41 | 4.69 | 3.87 | 4.12 |

## V. CONCLUSIONS

In this paper, we proposed EffiFusion-GAN, an efficient model for speech enhancement tasks. By incorporating depthwise separable convolutions, we effectively reduced the parameter count, significantly improving computational efficiency. Additionally, the model integrates a residual attention mechanism, enhancing feature extraction accuracy and improving the model's understanding of complex speech signals. Finally, the use of pruning techniques further optimized the model's inference speed and memory usage, ensuring high speech enhancement quality while reducing its complexity. Experimental results demonstrate that EffiFusion-GAN achieves comparable or even superior performance across various speech enhancement metrics while maintaining a smaller parameter footprint compared to other state-of-the-art methods. In the future, we aim to extend EffiFusion-GAN to more speech processing tasks and explore further optimization strategies to improve its performance.

## VI. ACKNOWLEDGEMENT

This research was supported by Horizon 2020 grant: Exchanges for SPEech ReseArch aNd TechnOlogies (ESPERANTO), Grant agreement ID: 101007666.

## REFERENCES

[1] A. Pandey and D. Wang, "TCNN: Temporal convolutional neural network for real-time speech enhancement in the time domain," in Proc. ICASSP, 2019, pp. 6875–6879.

[2] A. Defossez, G. Synnaeve, and Y. Adi, "Real time speech enhancement in the waveform domain," in Proc. Interspeech, 2020, pp. 3291–3295.

[3] E. Kim and H. Seo, "SE-Conformer: Time-domain speech enhancement using conformer." in Proc. Interspeech, 2021, pp. 2736–2740.

[4] K. Paliwal, K. Wojcicki, and B. Shannon, "The importance of phase in speech enhancement," Speech Communication, vol. 53, no. 4, pp. 465–494, 2011.

[5] M. Strauss and B. Edler, "A flow-based neural network for time domain speech enhancement," in *ICASSP 2021-2021 IEEE International Conference on Acoustics, Speech and Signal Processing (ICASSP)*, 2021, pp. 5754-5758.

[6] F. Chollet, "Xception: Deep learning with depthwise separable convolutions," in Proc. IEEE Conf. Comput. Vis. Pattern Recognit. (CVPR), 2017, pp. 1251–1258.

[7] M. Sandler, A. Howard, M. Zhu, A. Zhmoginov, & L. C. Chen, "Mobilenetv2: Inverted residuals and linear bottlenecks," in Proc. IEEE Conf. Comput. Vis. Pattern Recognit. (CVPR), 2018, pp. 4510-4520.

[8] X. Zhang, X. Zhou, M. Lin, & J. Sun, "Shufflenet: An extremely efficient convolutional neural network for mobile devices," in Proc. IEEE Conf. Comput. Vis. Pattern Recognit. (CVPR), 2018, pp. 6848-6856.

[9] R. Cao, S. Abdulatif, and B. Yang, "CMGAN: Conformer-based Metric GAN for Speech Enhancement," in Proc. Interspeech, 2022, pp. 936–940.

[10] Y. Ai and Z.-H. Ling, "Neural speech phase prediction based on parallel estimation architecture and anti-wrapping losses," in Proc. ICASSP, 2023.

[11] A. Pandey and D. Wang, "Densely connected neural network with dilated convolutions for real-time speech enhancement in the time domain," in Proc. ICASSP, 2020, pp. 6629–6633.

[12] K. He, X. Zhang, S. Ren, and J. Sun, "Delving deep into rectifiers: Surpassing human-level performance on imagenet classification," in Proc. ICCV, 2015, pp. 1026–1034.

[13] Lu Y X, Ai Y, Ling Z H. MP-SENet: A speech enhancement model with parallel denoising of magnitude and phase spectra[J]. arXiv preprint arXiv:2305.13686, 2023.

[14] S.-W. Fu, C. Yu, T.-A. Hsieh, P. Plantinga, M. Ravanelli, X. Lu, and Y. Tsao, "MetricGAN+: An improved version of MetricGAN for speech enhancement," in Proc. Interspeech, 2021, pp. 201–205.

[15] S. Pascual, A. Bonafonte, and J. Serra, "SEGAN: Speech enhancement generative adversarial network," in Proc. Interspeech, 2017, pp. 3642–3646.

[16] S.-W. Fu, C.-F. Liao, Y. Tsao, and S.-D. Lin, "MetricGAN: Generative adversarial networks based black-box metric scores optimization for speech enhancement," in Proc. ICML, 2019, pp. 2031–2041.

[17] F. Dang, H. Chen, and P. Zhang, "DPT-FSNet: Dual-path transformer based full-band and sub-band fusion network for speech enhancement," in Proc. ICASSP, 2022, pp. 6857–6861.

[18] D. Yin, C. Luo, Z. Xiong, and W. Zeng, "PHASEN: A phase-and harmonics-aware speech enhancement network," in Proc. AAAI, vol. 34, no. 05, 2020, pp. 9458–9465.

[19] C. Valentini-Botinhao, X. Wang, S. Takaki and J. Yamagishi, "Investigating RNN-based speech enhancement methods for noise-robust text-to-speech." in SSW, 2016, pp. 146–152.

[20] J. Thiemann, N. Ito, and E. Vincent, "The diverse environments multi-channel acoustic noise database (DEMAND): A database of multichannel environmental noise recordings," in Proc. ICA, vol. 19, no. 1, 2013, p. 035081.

[21] Drgas S. A Survey on Low-Latency DNN-Based Speech Enhancement[J]. Sensors, 2023, 23(3): 1380.